\documentclass[apj]{emulateapj}
\usepackage{natbib}
\usepackage{epstopdf}
\usepackage{CJK}
\usepackage{color}
\usepackage{pshan}
\usepackage{amsmath}

\newcommand{\beq}{\begin{equation}}
\newcommand{\eeq}{\end{equation}}
\newcommand{\bea}{\begin{eqnarray}}
\newcommand{\eea}{\end{eqnarray}}

\begin{document}
\begin{CJK}{UTF8}{mj}
\title{First Structure Formation under the Influence of Gas-Dark Matter Streaming Velocity and Density: Impact of the Baryons-trace-dark matter Approximation}

\author{Hyunbae Park (박현배)\altaffilmark{1}}
\author{Kyungjin Ahn\altaffilmark{2}}
\author{Naoki Yoshida\altaffilmark{1,3,4}}
\author{Shingo Hirano\altaffilmark{5}}
\altaffiltext{1}{Kavli IPMU (WPI), UTIAS, The University of Tokyo, Kashiwa, Chiba 277-8583, Japan; hyunbae.park@ipmu.jp}
\altaffiltext{2}{Department of Earth Sciences, Chosun University, Gwangju 61452, Republic of Korea; kjahn@chosun.ac.kr}
\altaffiltext{3}{Department of Physics, School of Science, The University of Tokyo, Bunkyo, Tokyo 113-0033, Japan}
\altaffiltext{4}{Research Center for the Early Universe, School of Science, The University of Tokyo, Bunkyo, Tokyo 113-0033, Japan}
\altaffiltext{5}{Department of Earth and Planetary Sciences, Faculty of Sciences, Kyushu University, Fukuoka, Fukuoka 819-0395, Japan}

\begin{abstract} 
The impact of the streaming between baryons and dark matter on the first structures has been actively explored by recent studies. We investigate how the key results are affected by two popular approximations. One is to implement the streaming by accounting for only the relative motion while assuming ``baryons trace dark matter" spatially at the initialization of simulation. This neglects the smoothing on the gas density taking place before the initialization. In our simulation initialized at $z_i=200$, it overestimates the gas density power spectrum by up to 40\% at $k\approx10^2~h~\mbox{Mpc}^{-1}$ at $z=20$. Halo mass ($M_h$) and baryonic fraction in halos ($f_{b,h}$) are also overestimated, but the relation between the two remains unchanged. The other approximation tested is to artificially amplify the density/velocity fluctuations in the cosmic mean density to simulate the first minihalos that form in overdense regions. This gives a head start to the halo growth while the subsequent growth is similar to that in the mean density. The growth in a true overdense region, on the other hand, is accelerated gradually in time. For example, raising $\sigma_8$ by 50\% effectively transforms $z\rightarrow\sqrt{1.5}z$ in the halo mass growth history while in 2-$\sigma$ overdensity, the growth is accelerated by a constant in redshift: $z\rightarrow{z+4.8}$. As a result, halos have grown more in the former than in the latter before $z\approx27$ and vice versa after. The $f_{b,h}$-$M_h$ relation is unchanged in those cases as well, suggesting that the Pop III formation rate for a given $M_h$ is insensitive to the tested approximations.
\end{abstract}


\section{Introduction}

Formation of the first stars (population III or ``Pop III'' star) is an important milestone in cosmic history, where the primordial density fluctuations from cosmic inflation \citep{1981PhRvD..23..347G,1982PhLB..108..389L} started collapsing ambient baryons into bound objects from $z\sim30$, which led to the production of ultraviolet radiation into space for the first time in the cosmic history \citep[e.g.,][]{2013RPPh...76k2901B,BARKANA2001125}.  According to the standard $\Lambda$CDM model of the structure formation, the structures began collapsing from small scales followed by their assembly into larger structures. Low-mass dark matter halos with $\sim 10^4$--$10^8 M_\odot$ (i.e., minihalos) are considered as the formation sites of the first collapsed objects. The details of the collapse involve highly nonlinear physics and are an active field of numerical astrophysics \citep[e.g.,][]{2003ApJ...592..645Y}. 

Recently, \cite{2010PhRvD..82h3520T} pointed out that the residual velocity fluctuations from the baryonic acoustic oscillation (BAO) resulted in a strong relative motion of typically $\sim 30 $ km/s between baryons and dark matter at cosmic recombination. This motion decayed in time, but it was strong enough to induce the streaming of gas through dark matter potential wells and thus make it more difficult for minihalos to grow their masses and accrete gas at the time of the first star formation. Subsequent numerical studies confirmed that the baryonic fraction in minihalos is highly suppressed by the streaming motion \citep{2011ApJ...736..147G,2012ApJ...760....4O,2013ApJ...763...27N,2013ApJ...771...81R,2016PhRvD..93b3518A}. Moreover, supersonic motion shock heats the gas, making cold gas even rarer inside halos \citep{2019MNRAS.484.3510S}. 

The global impact of the streaming motion on cosmic reionization is being actively explored. The beginning of the reionization is expected to be delayed \citep{2011MNRAS.412L..40M,2019ApJ...877L...5S} although the impact is considered to be limited at the late stage of reionization ($z\sim6$), which is driven by more massive ($\gtrsim 10^8 ~M_\odot$) atomic-cooling halos \citep{2011ApJ...730L...1S, 2014Natur.506..197F}. In semi-numerical models of star-formation and reionization, the streaming is considered to raise the minimum halo mass that can form Pop III stars \citep[e.g.,][]{2011ApJ...736..147G,2019PhRvD.100f3538M, 2020arXiv200111118V}. Also, the effect is expected to vary spatially because the streaming velocity is known to fluctuate at the BAO scale ($\sim 140~\rm{Mpc}$). It is an interesting possibility that large-scale fluctuations in the Pop III star-formation rate can leave an imprint on the spin temperature of atomic hydrogen \citep{McQuinn2012,2012Natur.487...70V,2019PhRvD.100f3538M}, which may be proved by upcoming 21cm surveys such as the Hydrogen Epoch of Reionization Array (HERA) and the Square Kilometre Array \citep[SKA:][]{2013ExA....36..235M,2014MNRAS.437L..36F, 2017PASP..129d5001D}.

There are also attempts to explain existing tensions between the standard cosmology and observation using the streaming motion. Regarding the mystery of high-redshift ($z\sim6$) supermassive black holes with $\sim 10^9~M_\odot$ \citep{2011Natur.474..616M,2015Natur.518..512W,2018ApJ...861L..14B}, several numerical studies have shown that the streaming can induce the formation of direct collapse black holes (DCBH) of $\sim 10^{5-6}~M_\odot$ at $z\sim 30$ \citep{2014MNRAS.439.1092T,2017Sci...357.1375H} to give a head start to the black hole growth although there are counter-arguments to this scenario \citep{2014MNRAS.440.2969L, 2014MNRAS.442L.100V}. Some studies have attempted to explain the formation mechanisms of missing satellites and globular clusters based on the fact that the streaming separates dark matter and baryons \citep{2013ApJ...768...70B,2014ApJ...791L...8N,2016MNRAS.460.1625P,2019ApJ...878L..23C}.

Given the increasing number of numerical simulation studies of the streaming motion, it is worth investigating the validity of approximations often made at the initialization of simulations. The first approximation to test is the assumption that {\it baryons trace dark matter} (BTD) at the initial conditions. Commonly used initial condition generators mostly assume the initial density/velocity field of baryonic matter is same as that of the dark matter at initialization. The actual amplitude of baryon density fluctuation is smaller than that of dark matter at $z\gtrsim 100$, but these two amplitudes are known to converge toward each other due to gravity before the first objects start forming. Thus, many numerical studies applied the streaming effect by simply adding a constant velocity to the baryon velocity field in the initial conditions, while using the same density field for both baryons and dark matter. This BTD assumption, however, is likely to break down when the streaming velocity shifts one component from the other. Also, the streaming effect should be stronger at higher redshift, but this approximation misses the effect taking place between cosmic recombination and the initialization of the simulation. To avoid this issue, one should either account for the effect in the density field at the initialization redshift or simply initialize the simulation at the recombination, as in \citet{2018MNRAS.474.2173H}.

Another approximation to test is to artificially increase $\sigma_8$, which will amplify the density fluctuation at all scales, to mimic an overdense patch of the universe. This method is frequently used to assimilate the biased formation of the first structures in dense regions of the universe
\citep[e.g.,][]{2011ApJ...736..147G,2011ApJ...730L...1S, 2017Sci...357.1375H,2018ApJ...855...17H,2019MNRAS.484.3510S}.
 In a different context, some early simulation works based on the first-year Wilkinson Microwave Anisotropy Probe (WMAP) results often used $\sigma_8=0.9$, which is higher than the currently known value \citep[e.g.,][]{2003ApJ...598...73Y}. We shall examine how the structure growth compares between such cases and truly overdense cases.

To provide self-consistent initial conditions with the streaming motion in overdensity, \cite{2016ApJ...830...68A} developed a quasi-linear perturbation theory of small-scale fluctuations under the influence of a large-scale overdensity and the streaming-velocity environment. \cite{2018ApJ...869...76A} then developed an initial condition generator, BCCOMICS\footnote{https://github.com/KJ-Ahn/BCCOMICS} (Baryon-Cold dark matter COsMological Initial Condition generator for Small-scales), which calculates the perturbation equations of \cite{2016ApJ...830...68A} and generates corresponding three-dimensional initial conditions of dark matter and baryons. BCCOMICS treats a given overdense (underdense) patch as a separate universe with positive (negative) curvature and provides a set of ``local cosmology parameters'' to account for the local expansion rate being different from the mean cosmic expansion rate. \cite{2018ApJ...869...76A} used BCCOMICS to generate a suite of initial conditions for varying streaming-velocity and density environments and then performed $N$-body and hydrodynamic simulations to explore the cosmic variance of high-redshift structure formation. 

This study is a continuation of the efforts by \cite{2016ApJ...830...68A} and \cite{2018ApJ...869...76A} to explore the dual impact of the streaming motion and overdensity with correctly generated initial conditions, and an extension of these works to compare the self-consistent approach quantitatively to the two common approximations used in generating initial conditions. Therefore, this work partially revisits the work of \cite{2012ApJ...760....4O}, which tested the baryons-trace-dark matter assumption by providing more results on key statistics of the Pop III star formation. 

The paper is organized as follows. In Section~\ref{sec:method}, we introduce our numerical methods used in this study. In Section~\ref{sec:results}, we show our results. In Section~\ref{sec:discussion}, we summarize our results and give conclusions. For the rest of this paper, we assume $\Lambda$CDM cosmology consistent with the WMAP 9-year results \citep{2013ApJS..208...19H}: $\Omega_{m,0}=0.276$, $\Omega_{b,0}=0.045$, $h=0.703$, $\sigma_8=0.8$ and $n_s=0.961$. 

\section{Methodology} \label{sec:method}

\begin{table*}[]
\begin{center} 
\caption{Parameters for the hydrodynamic simulations used in this study}
\label{tab:sims}
\begin{tabular}{cccccccc}
\hline
Label & $L_{\rm box}$ {[}$h^{-1}$ Mpc{]} & $m_{\rm dm}$ {[}$h^{-1}$ $M_\odot${]} & $m_{\rm gas}$ {[}$h^{-1}$ $M_\odot${]} & $V_{cb,1000}$ {[}km/s{]} & $\Delta_{z=200}$ & Baryons trace dark matter & $\sigma_8$ \\ \hline
d0v0 & 1 & $6.80\times 10^2 $ & $1.32 \times 10^2 $ & 0 & 0 & No & 0.8 \\
d0v2 & 1 & $6.80\times 10^2 $ & $1.32 \times 10^2 $ & 56 & 0 & No & 0.8 \\
d0v2L & 4 & $4.35\times 10^4 $ & $8.45\times 10^3 $ & 56 & 0 & No & 0.8 \\
d2v2 & 1 & $7.02\times 10^2 $ & $1.35 \times 10^2 $ & 56 & $3.14\times 10^{-2}$ & No & 0.8 \\
d0v2\_BTD & 1 & $6.80\times 10^2 $ & $1.32 \times 10^2 $ & 56 & 0 & Yes & 0.8 \\
d0v2L\_BTD & 4 & $4.35\times 10^4 $ & $8.45\times 10^3 $ & 56 & 0 & Yes & 0.8 \\
d0v2\_IS & 1 & $6.80\times 10^2 $ & $1.32 \times 10^2 $ & 56 & $0$ & No & 1.2 \\ \hline
\end{tabular}
\end{center} 
\end{table*}

   \begin{figure*}
  \begin{center}
    \includegraphics[scale=0.38]{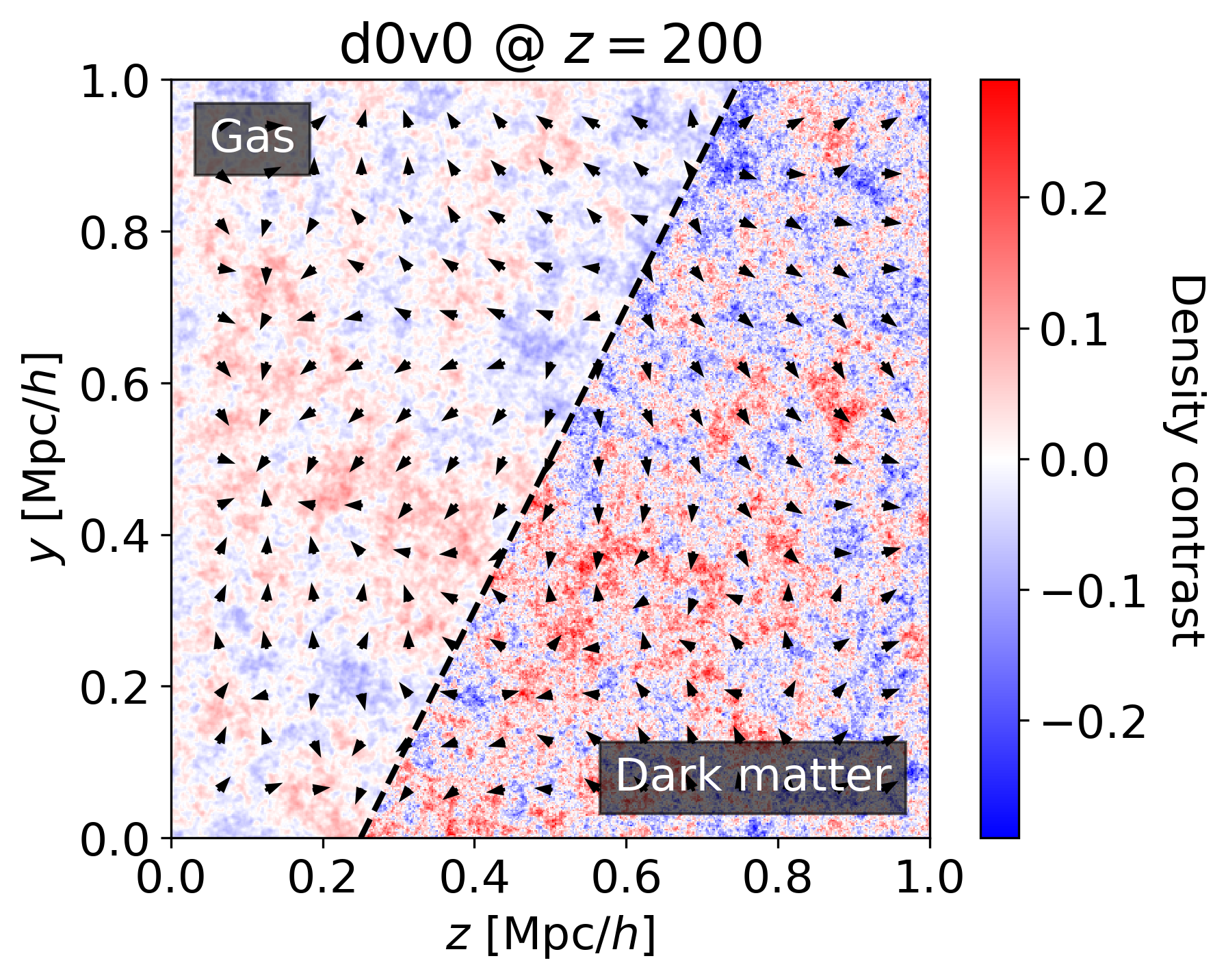}
    \includegraphics[scale=0.38]{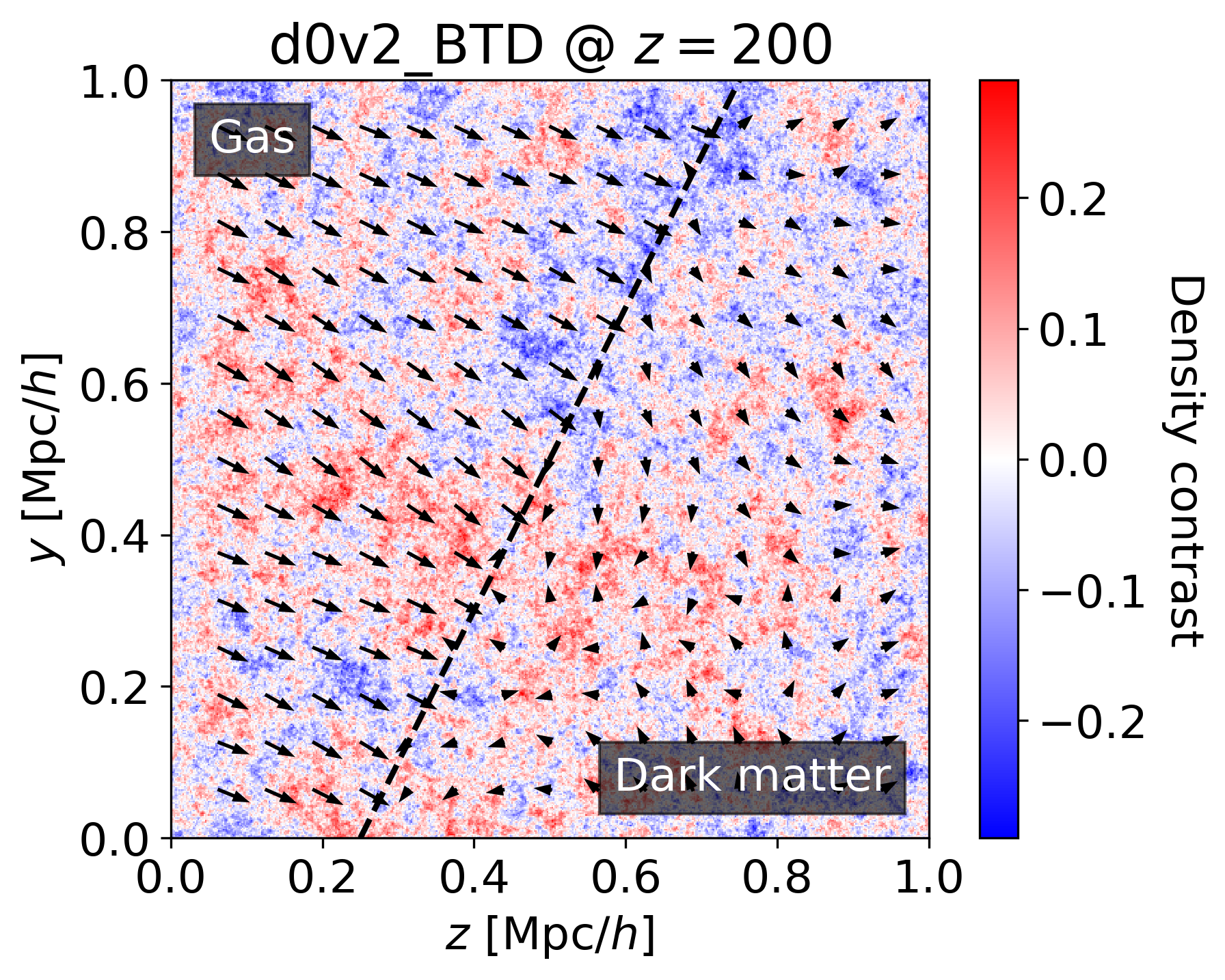}
    \includegraphics[scale=0.38]{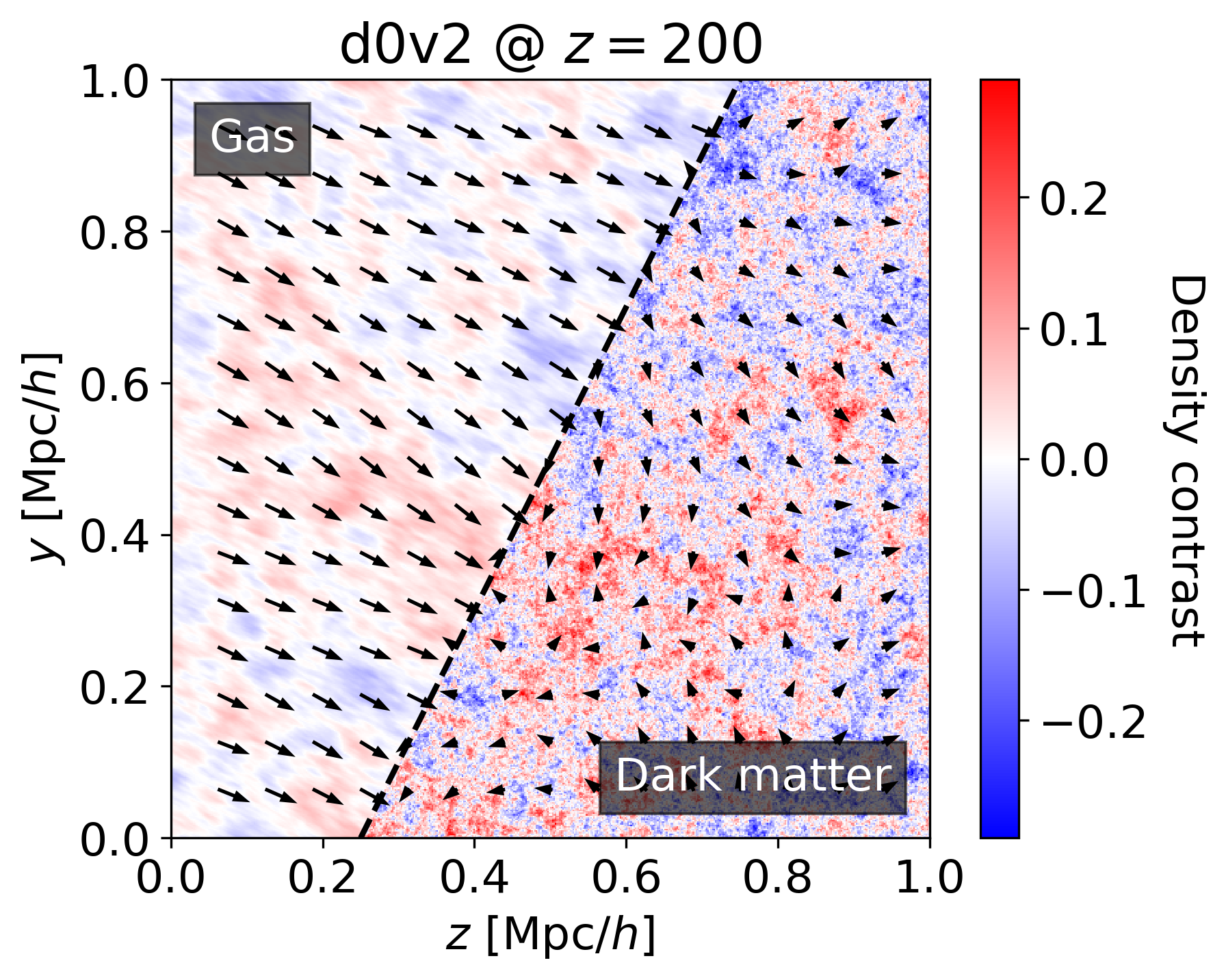}
  \caption{Initial gas density/velocity field of the simulation at $z=200$ visualized for d0v0 (left), d0v2\_BTD (middle), and d0v2 (right). Density and velocity are described by color contours and arrows, respectively. The density field is color-coded so that over/underdensity is shown in red/blue. }
  \label{fig:IC}
  \end{center}
\end{figure*}

\subsection{Basics of Streaming Motion} \label{sec:basics}

In the absence of the streaming velocity and non-gravitational baryonic physics, the perturbation equation for the overdensity $\delta$ and the peculiar velocity $v$ of matter is given by
 \bea \label{eq:stream0}
 \frac{\partial \delta}{\partial t} &=& -\theta \nonumber \\
 \frac{\partial \theta}{\partial t} &=& - \frac{3H^2}{2} \Omega_m \delta - 2H\theta,
 \eea
where $\theta\equiv a^{-1} \nabla\cdot\bold{v}$ with the scale factor $a$ and the gradient in the comoving frame $\nabla$, $H$ is the Hubble parameter, and $\Omega_m$ is the cosmic matter fraction of the universe at given cosmic time $t$. Common initial condition generators use the solution from the above equation to set the density and velocity fluctuation amplitudes of both baryons and dark matter.
  
The perturbation equation in the presence of baryon-dark matter streaming velocity ($\bold{V}_{cb}=-\bold{V}_{bc}\equiv{\bold{V}}_{c}-{\bold{V}}_{b}$ with the average peculiar velocities of CDM ${\bold V}_{c}$ and baryon ${\bold V}_{b}$ inside a patch) was first derived by  \citet[see Eq.~6 of their work]{2010PhRvD..82h3520T}.  \cite{2012ApJ...760....4O} used their initial condition generator, Cosmological Initial Conditions for AMR and smoothed particle hydrodynamics (SPH) Simulations (CICsASS), to generate initial conditions from the solution of the equation. Then, \cite{2016ApJ...830...68A} improved on the equation for non-zero overdensity ($\Delta$) as well accommodating the dual impact of $\bold{V}_{cb}$ and $\Delta$, because the original perturbation equation by \citet{2010PhRvD..82h3520T} does not implement the non-zero overdensity environment. \cite{2018ApJ...869...76A} developed the initial condition generator BCCOMICS based on \cite{2016ApJ...830...68A}. Their perturbation equation for the Fourier modes of density contrast ($\delta_b$ for baryons \& $\delta_c$ for dark matter), velocity divergence ($\theta_b$ for baryons \& $\theta_c$ for dark matter) and baryonic temperature fluctuations ($\delta_T \equiv (T-\bar{T})/{\bar{T}}$, where $T$ is local baryon temperature) reads
\bea \label{eq:stream}
 \frac{\partial \delta_c}{\partial t} &=& -(1+\Delta_c)\theta_c - \Theta_c \delta_c \nonumber  \\
 \frac{\partial \theta_c}{\partial t} &=& - \frac{3H^2}{2} \Omega_m (f_c \delta_c + f_b \delta_b ) - 2H\theta_c \nonumber  \\
 \frac{\partial \delta_b}{\partial t} &=& -ia^{-1}\bold{V}_{bc}\cdot \bold{k} \delta_b  - (1+\Delta_b) \theta_b - \Theta_b \delta_b \nonumber  \\
 \frac{\partial \theta_b}{\partial t} &=& -ia^{-1}\bold{V}_{bc}\cdot \bold{k} \theta_b  - \frac{3H^2}{2} \Omega_m (f_c \delta_c + f_b \delta_b ) - 2H\theta_b \nonumber  \\
 &&+a^{-2}\frac{k_B \bar{T}}{\mu m_H} k^2 \{ (1+\Delta_b)\delta_T  + (1+\Delta_T)\delta_b \} \nonumber  \\
 \frac{\partial  \delta_T}{\partial t} &=& \frac{2}{3} 
 \left\{ 
   \frac{\partial \delta_b}{\partial t}  +
  \frac{\partial \Delta_b}{\partial t}  (\delta_T-\delta_b) +
  \frac{\partial \delta_b}{\partial t}  (\Delta_T-\Delta_b)
 \right\} \nonumber \\
 &&- \frac{x_e(t)}{t_\gamma} a^{-4} \frac{\bar{T}_\gamma}{\bar{T}}\delta_T,
\eea
in the CDM-rest frame (${\bold V}_{c}=0$).
Here, $f_b=\Omega_b/\Omega_m$ and $f_c=(\Omega_m-\Omega_b)/\Omega_m$ are the global baryon and dark matter fraction in matter, respectively, $x_e$ is the global ionized fraction, $t_{\gamma}=1.17\times 10^{12}~{\rm yrs}$, and ${\bar T}_{\gamma}=2.725(1+z)~{\rm K}$ is the mean temperature of the cosmic microwave background at redshift $z$. The bulk quantities of a patch $\Delta_b$, $\Delta_c$, $\Theta_b (\equiv a^{-1}\nabla\cdot{\bf V}_b)$, $\Theta_c (\equiv a^{-1}\nabla\cdot{\bf V}_c)$, and $\Delta_T$ denote the overdensity of baryons, the overdensity of dark matter, the divergence of ${\bf V}_b$, the divergence of ${\bf V}_c$, and the baryon temperature fluctuation, respectively, and their values in Fourier space are identical to the real-space values. Due to the linearity of any perturbative quantities and the smallness of pressure terms at large scales, these bulk quantities satisfy the following linearized equation: 
 \bea \label{eq:bulk}
 \frac{\partial \Delta_c}{\partial t} &=& -\Theta_c , \nonumber \\
 \frac{\partial \Theta_c}{\partial t} &=& - \frac{3H^2}{2} \Omega_m (f_c \Delta_c + f_b \Delta_b) - 2H\Theta_c , \nonumber \\
 \frac{\partial \Delta_b}{\partial t} &=& -\Theta_b ,\nonumber \\
 \frac{\partial \Theta_b}{\partial t} &=& - \frac{3H^2}{2} \Omega_m (f_c \Delta_c + f_b \Delta_b) - 2H\Theta_b .
 \eea
$\Delta_b$ and $\Delta_c$ defined at the length scale $4~h^{-1}{\rm Mpc}$ are tightly correlated at $z\lesssim 200$ and almost uncorrelated at $z\sim 1000$ \citep{2016ApJ...830...68A,2018ApJ...869...76A}. In this work, we shall run several simulations with $\Delta_c=0.0323$ and $\Delta_b=0.027$ at $z=200$, which correspond to 2-$\sigma$ overdensity for a $4~h^{-1}{\rm Mpc} $ box at that redshift.

The streaming velocity fluctuates spatially at the BAO scale ($\sim 140$ Mpc). Thus, the streaming velocity $\bold{V}_{cb}$ can be treated as a constant drift within $10$ Mpc. $|\bold{V}_{cb}|$ follows a Boltzmann distribution with the standard deviation of $\sigma=$ 28 km/s at $z=1000$, which decays as $(1+z)$ with cosmic expansion. We shall take its value at $z=1000$, $V_{cb,1000}\equiv |\bold{V}_{cb}(z=1000)|)$, as the reference value.

\subsection{Simulation Setup} \label{sec:simlist}

\subsubsection{Parameter Choice}

The list of the simulation parameter choices for the simulations in this work is given in Table~\ref{tab:sims}. The fiducial case, d0v0, has the cosmic mean density and zero streaming velocity in a $1~h^{-1}{\rm Mpc}$ box. Several cases are run with a streaming velocity of $56[z/1000]$ km/s, which is twice the root-mean-square of the streaming velocity distribution.  We run a streaming case in the cosmic mean density (d0v2) and in the 2-$\sigma$ overdensity (d2v2; $\Delta=3.14\times10^{-2}$). We also run one simulation with the 2-$\sigma$ streaming velocity and the cosmic mean density in a bigger box of $4~h^{-1}{\rm Mpc}$ to obtain statistics of higher-mass halos (d0v2L) that cannot be captured in a $1~h^{-1}{\rm Mpc}$ box.

We make two cases with the two above-mentioned approximations. In d0v2\_BTD, we apply the BTD assumption in the initial conditions by assigning the same density field to both baryons and dark matter. In this case, the amplitude of the density/velocity fluctuations is given by Equation~(\ref{eq:stream0}) and a constant streaming velocity is added to the baryon velocity field. We also run a $4~h^{-1}{\rm Mpc}$ box simulation with the same setup (d0v2L\_BTD). In d0v2\_IS, we artificially boost the normalization of the initial density power spectrum in d0v2 by raising $\sigma_8$ from $0.8$ to $1.2$ as done in some previous works to simulate overdense regions.

\subsubsection{Initial Conditions}

   \begin{figure}
  \begin{center}
    \includegraphics[scale=0.5]{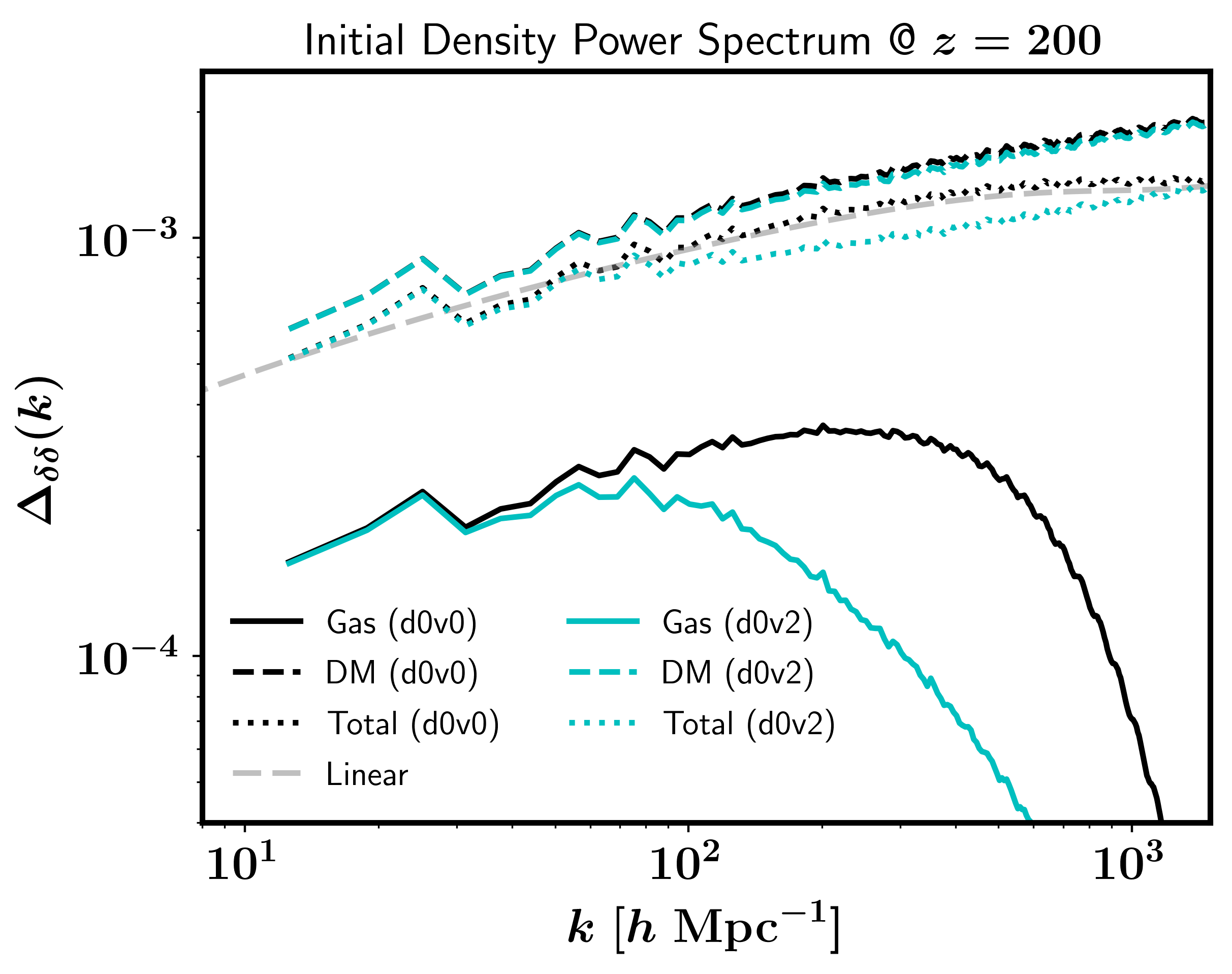}
  \caption{Density power spectrum of gas (solid lines), dark matter (dashed lines), and the total matter (dotted lines) density field for the case with (d0v2; cyan) and without the streaming motion (d0v0; black) at $z=200$. The linear matter-density power spectrum is shown as a gray dashed line for a reference.}
  \label{fig:ICPS}
  \end{center}
\end{figure}

 \begin{figure}
  \begin{center}
    \includegraphics[scale=0.5]{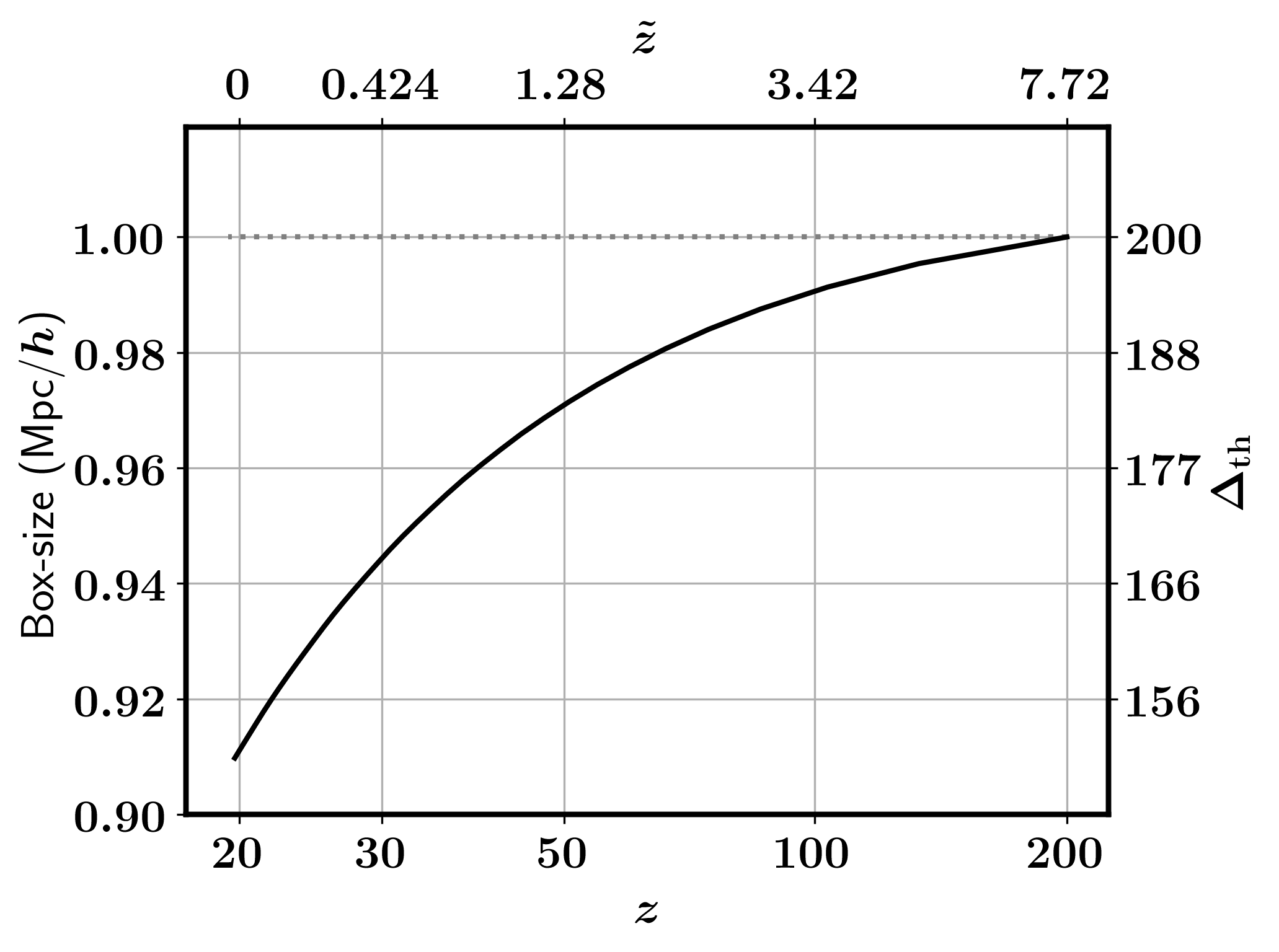}
  \caption{Comoving box-size of d2v2 as a function of redshift (black solid line). The upper $x$-axis shows the corresponding local redshift $\tilde{z}$ used internally in the simulation. The right-hand-side y-axis shows the threshold density for the halo radius in units of mean density of the simulation $(\Delta_{\rm th})$.}
  \label{fig:d2fac}
  \end{center}
\end{figure}

The initial conditions are generated for five cases in $1~h^{-1}{\rm Mpc}$ boxes and two cases in $4~h^{-1}{\rm Mpc}$ boxes at $z=200$. Simulations with the same box size are initialized with the same set of random phases to exclude the cosmic variance effect in the comparison.

In Figure~\ref{fig:IC}, we visualize the initial gas/dark matter density fields of d0v0, d0v2\_BTD and d0v2 at $z=200$ generated by BCCOMICS. The similarity in large-scale structures is due to the same random phases used for all three cases. At $z=200$, the density fluctuation amplitude of gas is smaller than that of dark matter in d0v0 and d0v2. The gas density fluctuation amplitude in d0v2\_BTD is highly overestimated due to the BTD approximation. Also, the smoothing effect from the streaming motion is not present in d0v2\_BTD. 

The power spectra of the initial gas, dark matter, and the total matter density are shown in Figure~\ref{fig:ICPS} for d0v0 and d0v2. In both cases, the dark matter density power spectrum is higher than the linear power spectrum while the gas density power spectrum is lower. The total matter power in d0v0 agrees exactly with the linear spectrum. In the case of d0v2\_BTD, both gas and dark matter density power spectra are the same as the linear power spectrum. The gas density power spectra of d0v0 and d0v2 agree up to $k=10^2~h~{\rm Mpc}^{-1}$, above which the spectrum of d0v2 falls off due to the suppression from the streaming motion.

We note that we use separate transfer functions for the gas and dark matter components by definition, except for the BTD cases, which used the transfer function of the total matter. Generating gas and dark matter density fields from the same transfer function is known to cause mild systematic effects that were explored in detail by  \citet{2003ApJ...592..645Y} and \citet{2020arXiv200200015B}. Our simulation setup is more relevant to the results of \citet{2003ApJ...592..645Y}, which are from $4~h^{-1}~{\rm Mpc}$ boxes at $z\ge30$. Their work showed that the BTD approximation fails to reproduce the offset between the gas and dark matter density power spectra, which persists to $z=30$ at $10-20$\% level at $k\sim 1-30 ~h~{\rm Mpc}^{-1}$.

  \begin{figure*}
  \begin{center}
  \includegraphics[scale=0.38]{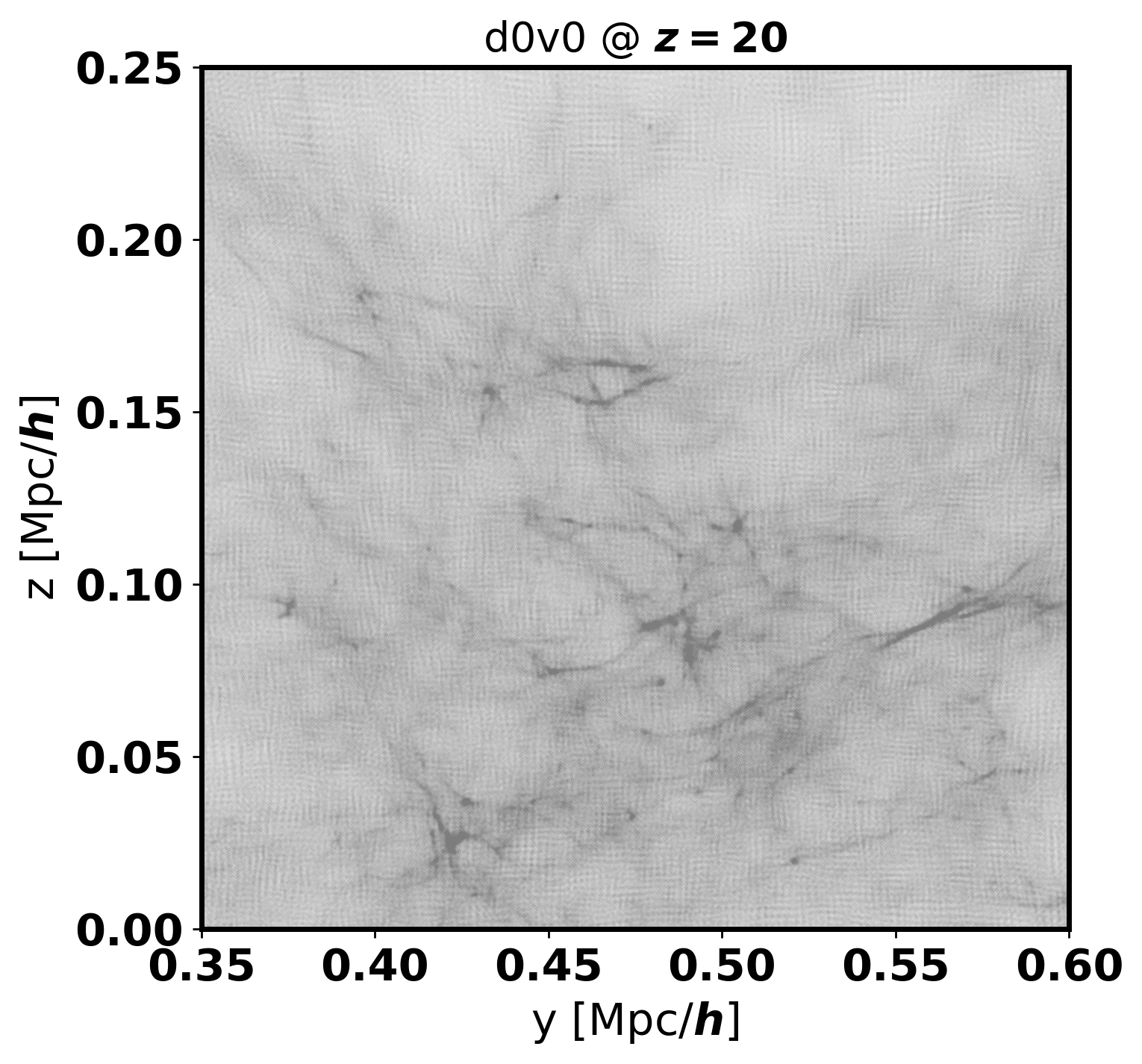}
  \includegraphics[scale=0.38]{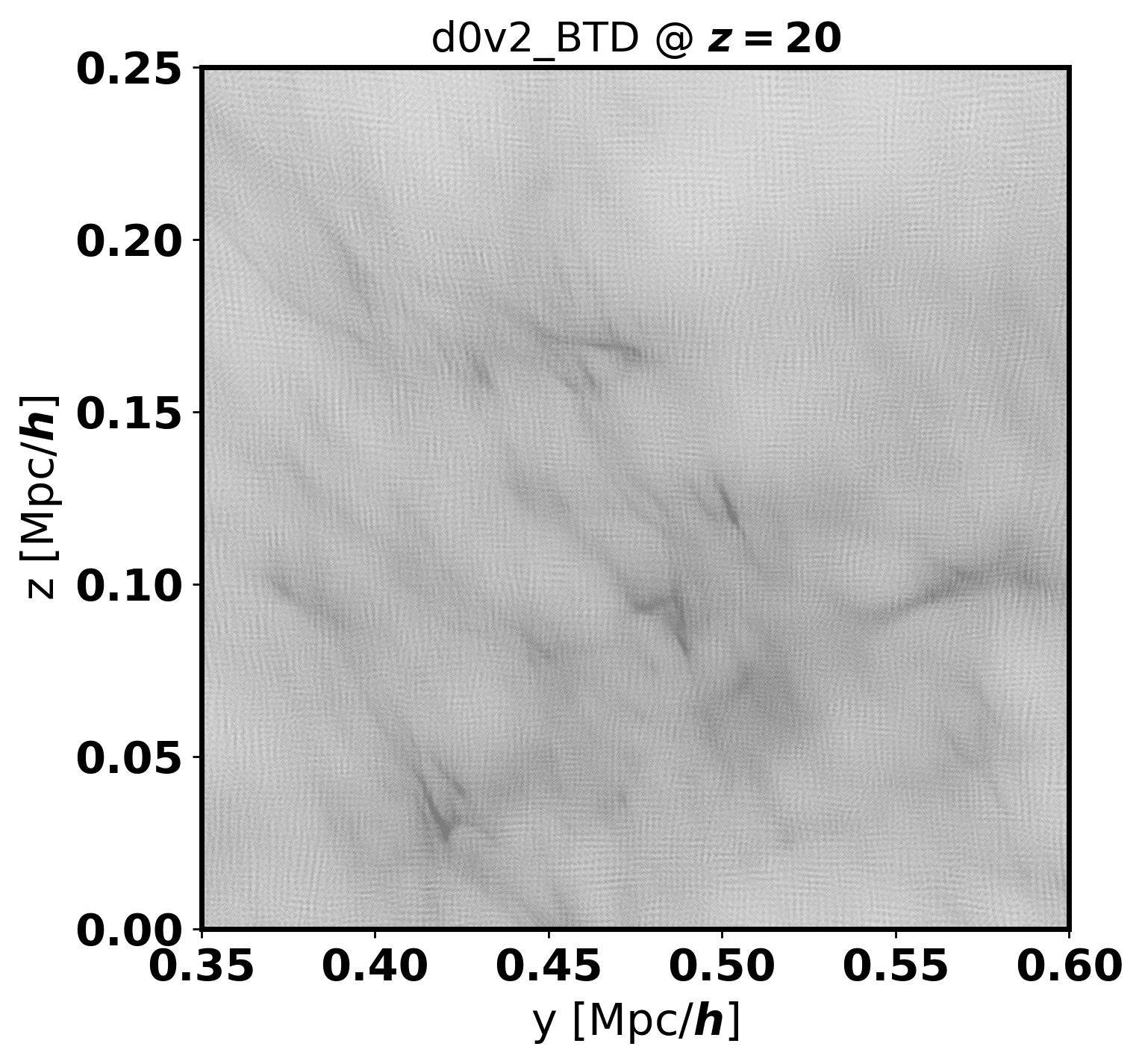}
  \includegraphics[scale=0.38]{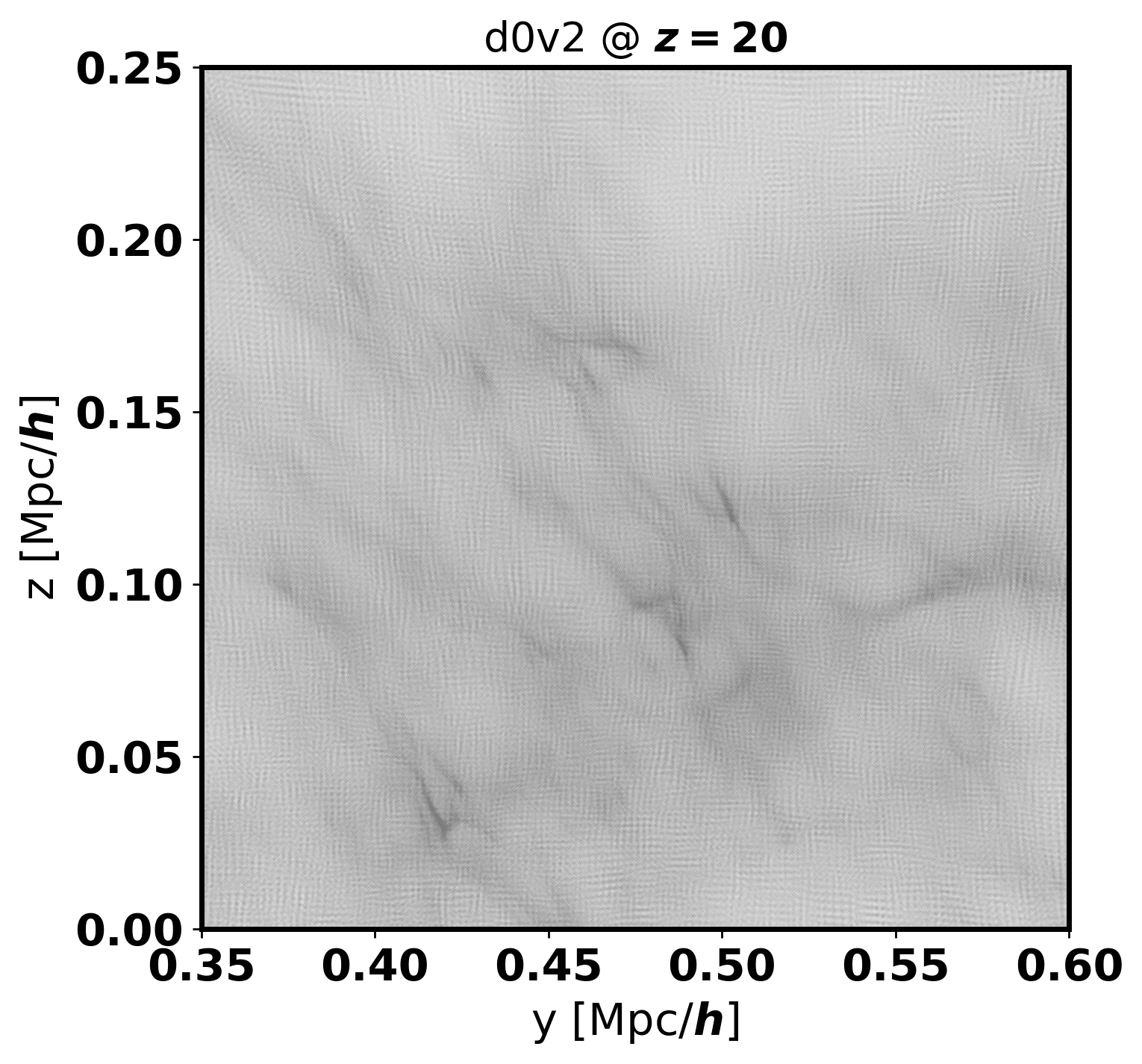}
  \caption{Gas particle distribution of d0v0 (left), d0v2\_BTD (middle), and d0v2 (right) at $z=20$. Density structures are depicted in dark gray.  A thin slice of a $0.25 \times 0.25~(h^{-1}{\rm Mpc})^2$ region, which has relatively more-grown structures is chosen for comparison.}
  \label{fig:ptl_map}
  \end{center}
\end{figure*}

\subsubsection{Hydrodynamics Solver}
For the hydrodynamics solver, we adopt the SPH code GADGET-2 \citep{2001MNRAS.328..726S,2005MNRAS.364.1105S} to follow the structure formation from $z=200$ to $\sim 15$. BCCOMICS creates the initial conditions in the ENZO \citep{Bryan2014} simulation format. We then convert those initial conditions into the GADGET format. We initialize the gas and dark matter particles on two separate grids that are offset by half the average particle distance along all of the three ($x$, $y$, $z$) coordinates.

\subsection{Simulating the Overdense Region}

The first collapsed objects likely appeared in overdense patches of the universe. In this work, we simulate one case in an overdense region (d2v2) to study the accelerated growth of structure in true overdensity.

Many numerical studies rely on a multi-resolution adaptive-refinement scheme \citep[e.g., MUSIC;][]{2011MNRAS.415.2101H} to start from a large box with the cosmic mean density and zoom into a denser subregion where structures develop earlier than in other parts of the volume. 

In this work, we present a complementary method, referred to as the ``separate universe'' approach, which starts from initial conditions of overdense volume using solutions of Equation~(\ref{eq:stream}) for a non-zero local overdensity $\Delta$. An advantage of this method is that one can easily create extremely rare density peaks, only a few of which appear in a ${\rm Gpc}^3$ volume. Creating such an extremely overdense patch would require an excessive number of refinements with the adaptive-resolution scheme. 

In an overdense region, the cosmic expansion rate is locally slower than the global rate. We capture this effect by modifying the cosmology parameters in the simulation setup. BCCOMICS provides the {\it local} cosmology parameters for volumes with non-zero overdensity. In an overdense volume, cosmology parameters of a closed universe are used to describe the expansion rate. 
The derivation of the local cosmology parameters is given in Section 3 of \cite{2018ApJ...869...76A}. A detailed description of this method can also be found from \cite{2005ApJ...634..728S} for cases without the streaming velocity.

Our method does not capture higher-order gravitational effects like the shear and tidal forces from missing large-scale structures. However, such effects should be negligible at the time of first structure formation. A similar approach was used by \cite{2004ApJ...605....1G} in their simulations of underdense regions to study cosmic voids. 

In volumes with non-zero overdensity, the redshift evolves differently from the global value due to the modified expansion rate. Thus, one must keep track of the relation between the true and the {\it local} redshift in the simulation. 
We define the local redshift $\tilde{z}$ so that it becomes zero at the end of the simulation: the simulations with non-zero overdensity end at the global redshift of $z=20$ in this study\footnote{ One may prefer to choose a later epoch to set $\tilde{z}=0$. We, however, note that high-density peaks may turn around and contract before $z=0$. For example, the d2v2 patch turns around at $z=4.6$ and even collapses to a point at $z=2.5$, which makes it impossible to perform the simulation until $z=0$. In this study, we therefore set the end of the simulation as the ``local present'' as in \cite{2018ApJ...869...76A}.}. 
In that case, the initial value of the local redshift in d2v2 is $\tilde{z}_i=7.72$ while the true initial redshift is $z_i=200$. In Figure~\ref{fig:d2fac}, we show how $\tilde{z}$ and the box size in the global comoving scale evolve in time. The local cosmological parameters of d2v2 are $\tilde\Omega_{\Lambda,0}=3.58\times10^{-4}$, $\tilde\Omega_{m,0}=1.43$, $\tilde\Omega_{b,0}=0.27$ and  $\tilde{h}=31.6$, where tildes are used to denote that the parameters are the local values. Note that the ``present'' values denoted by the subscript ``0'' of these cosmological parameters are evaluated at $\tilde{z}=0$.

It is worth noting that the simulation volume expands by a factor of $\tilde{z}_i+1=8.72$ between $z_i=200$ and $z_f=20$ while the universe globally expands by a factor of $[200+1]/[20+1]=9.57$. As a result, the comoving box-size of the overdense simulation shrinks by $\sim 9\%$ between $z_i$ and $z_f$ (see also Fig.~\ref{fig:d2fac}) and the mean density of the simulation is increased by $[9.57/8.72]^3=1.32$.

  \begin{figure*}
  \begin{center}
  \includegraphics[scale=0.5]{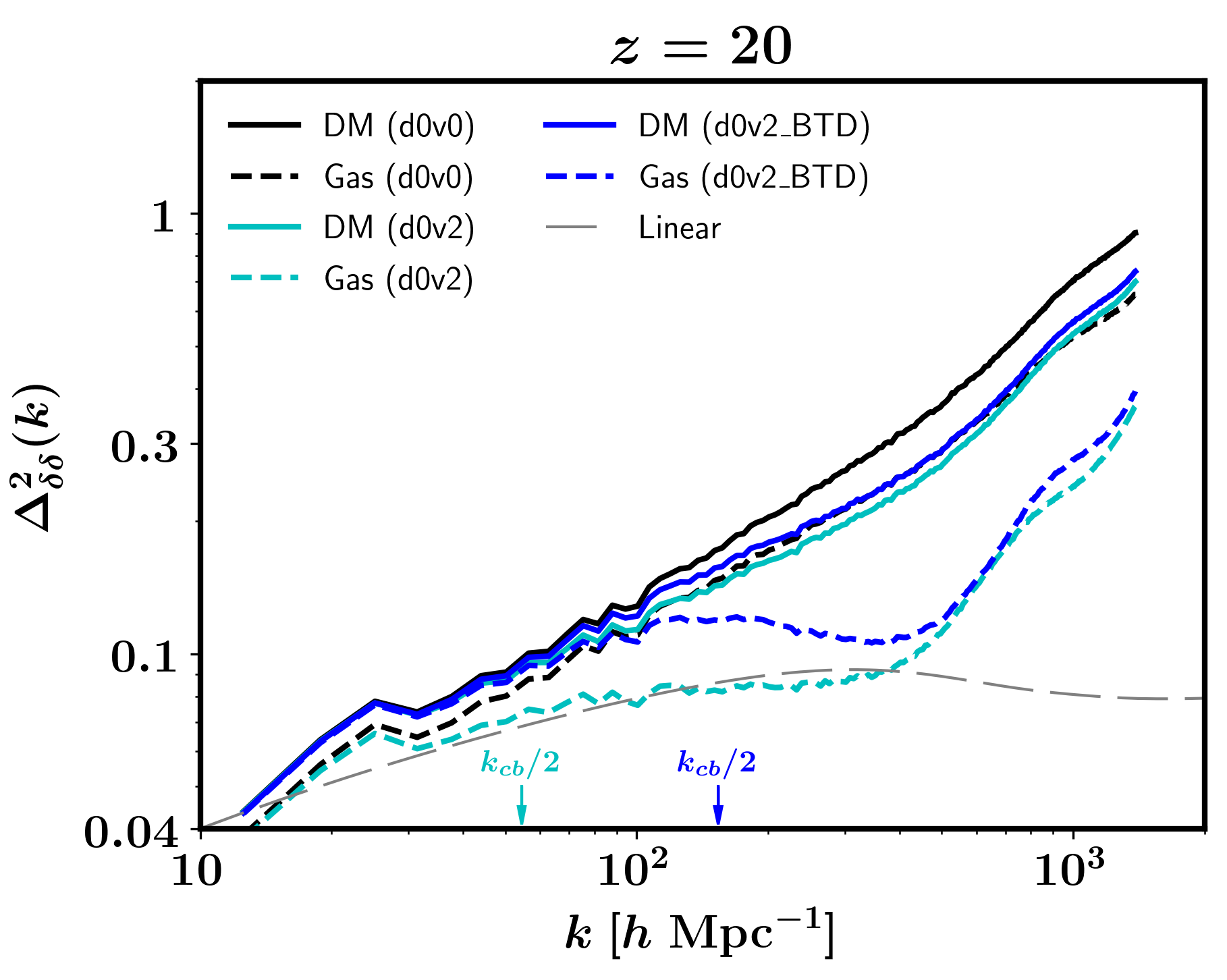}
  \includegraphics[scale=0.5]{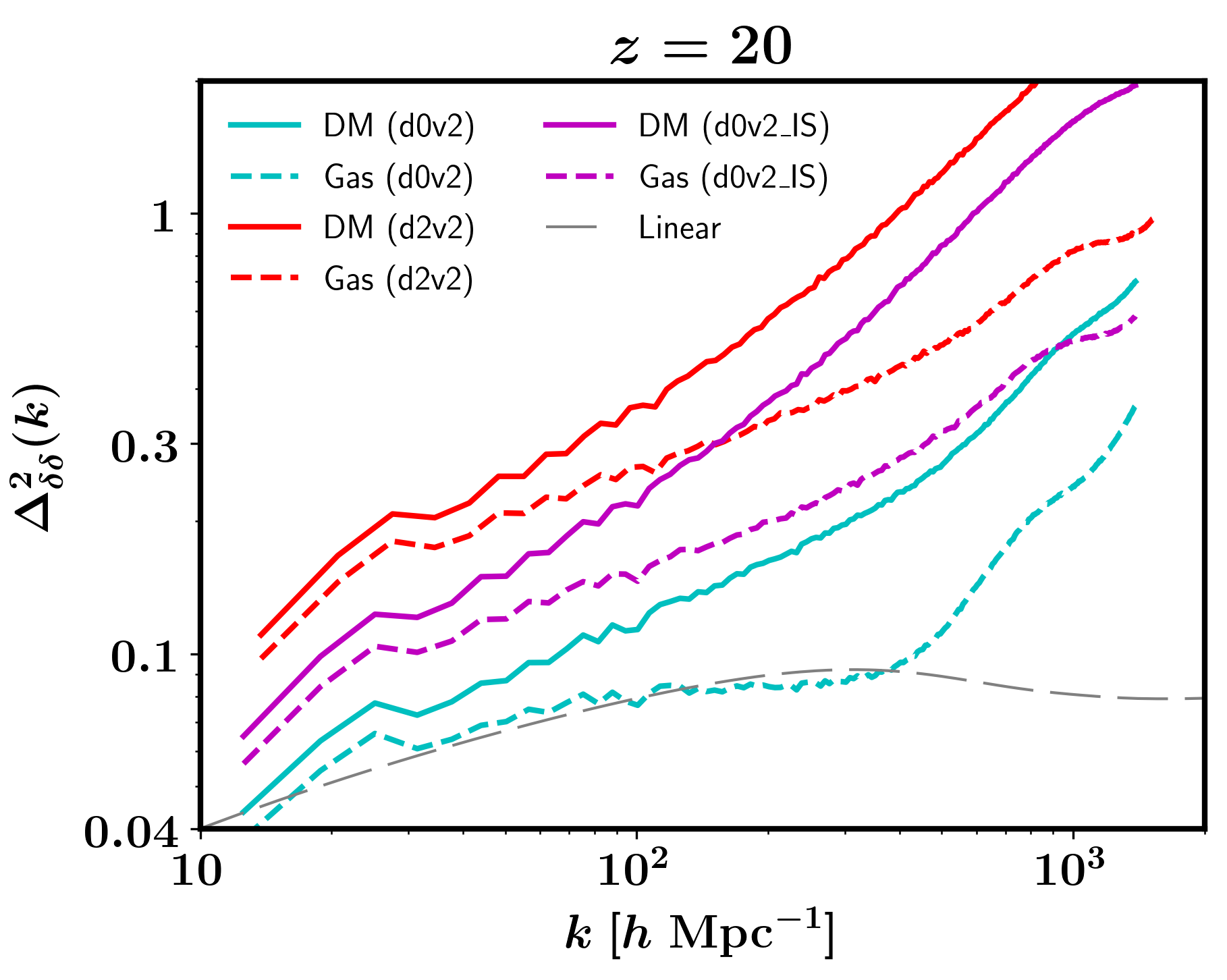}
  \caption{Dimensionless power spectrum of gas and dark matter density at $z=20$ plotted as solid and dashed lines, respectively. The left panel compares the results of d0v0 (black), d0v2 (cyan), and d0v2\_BTD (blue). Blue and cyan arrows denote $k_{cb}$ for d0v2\_BTD and d0v2, respectively. The right panel compares the results of d0v2 (cyan), d2v2 (red), and d0v2\_IS (magenta). To account for the overdensity in d2v2, we multiply the square of the overdensity factor $\Delta^2(z=20)=1.74$ to the power spectrum. The gray dashed line shows the linear density power spectrum for reference.}
  \label{fig:PS}
  \end{center}
\end{figure*}

\subsection{Halo Identification}

We use the publicly available version of Amiga Halo Finder\footnote{http://popia.ft.uam.es/AHF/Download.html} \citep[AHF;][]{2004MNRAS.351..399G,2009ApJS..182..608K} to identify halos from the simulation output at $z=20$. AHF outputs a list of gas, dark matter, and total mass of identified halos. These quantities are used for obtaining the halo mass function and baryonic fraction in halos for a given halo mass, which is considered highly relevant to the Pop III star formation rate.

As usual, the virial radius of a halo is chosen to make the mean density within halo $\Delta_{\rm th}=200$ times the cosmic mean. In the overdense case, the mean density of the simulation box increases when compared to the cosmic mean. We thus compensate for the overdensity by rescaling the density threshold parameter $\Delta_{\rm th}$, which is in units of the mean density of the simulation box as described in Figure~\ref{fig:d2fac}. For example, the virial radius of a halo in the overdense simulation at $z=20$ is defined by $\Delta_{\rm th}=200/1.32=152$ times the mean density of the simulation. The details of this mapping process are described in \cite{2018ApJ...869...76A}.

\section{Results} \label{sec:results}

 \begin{figure*}
  \begin{center}
    \includegraphics[scale=0.5]{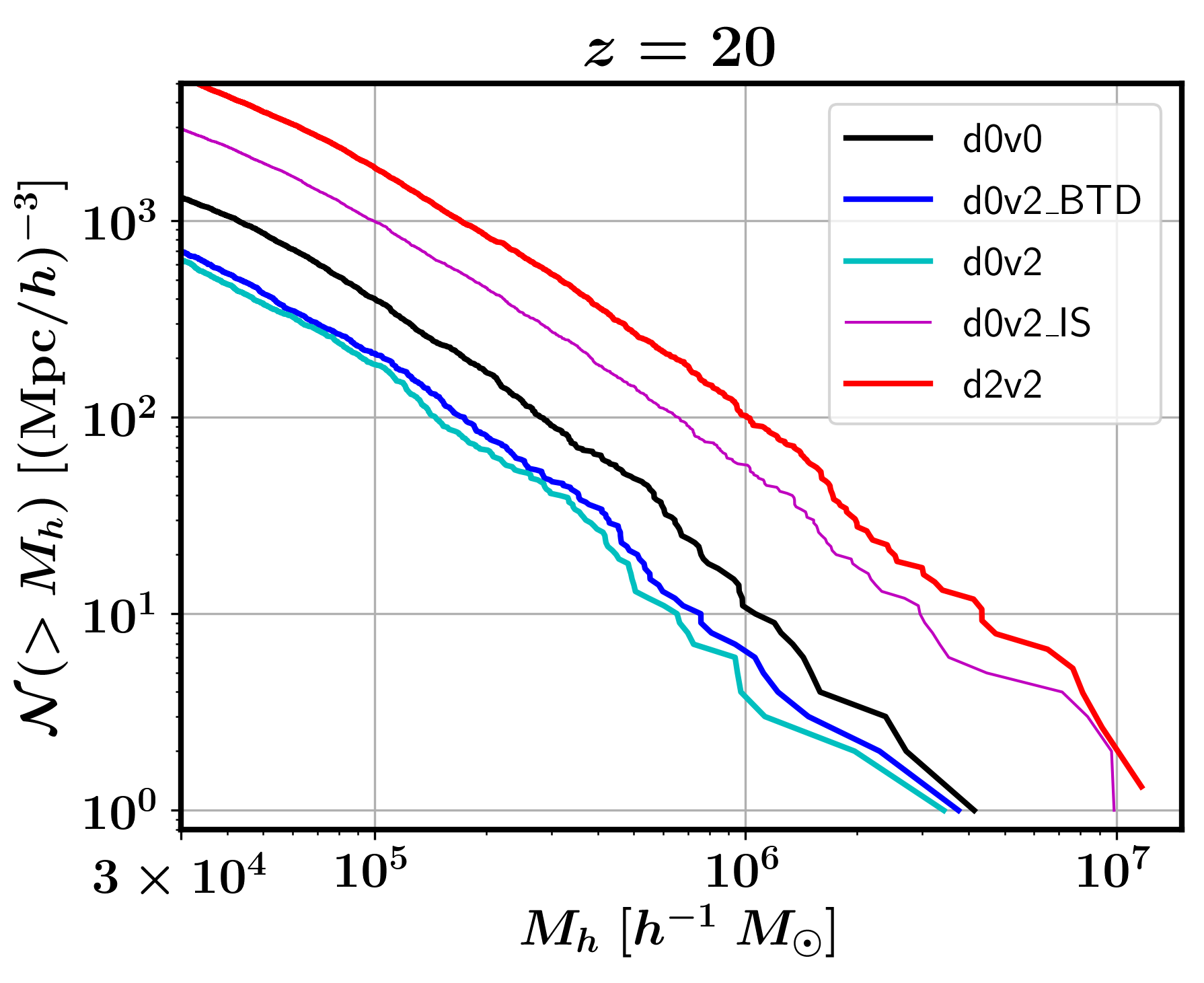}
    \includegraphics[scale=0.5]{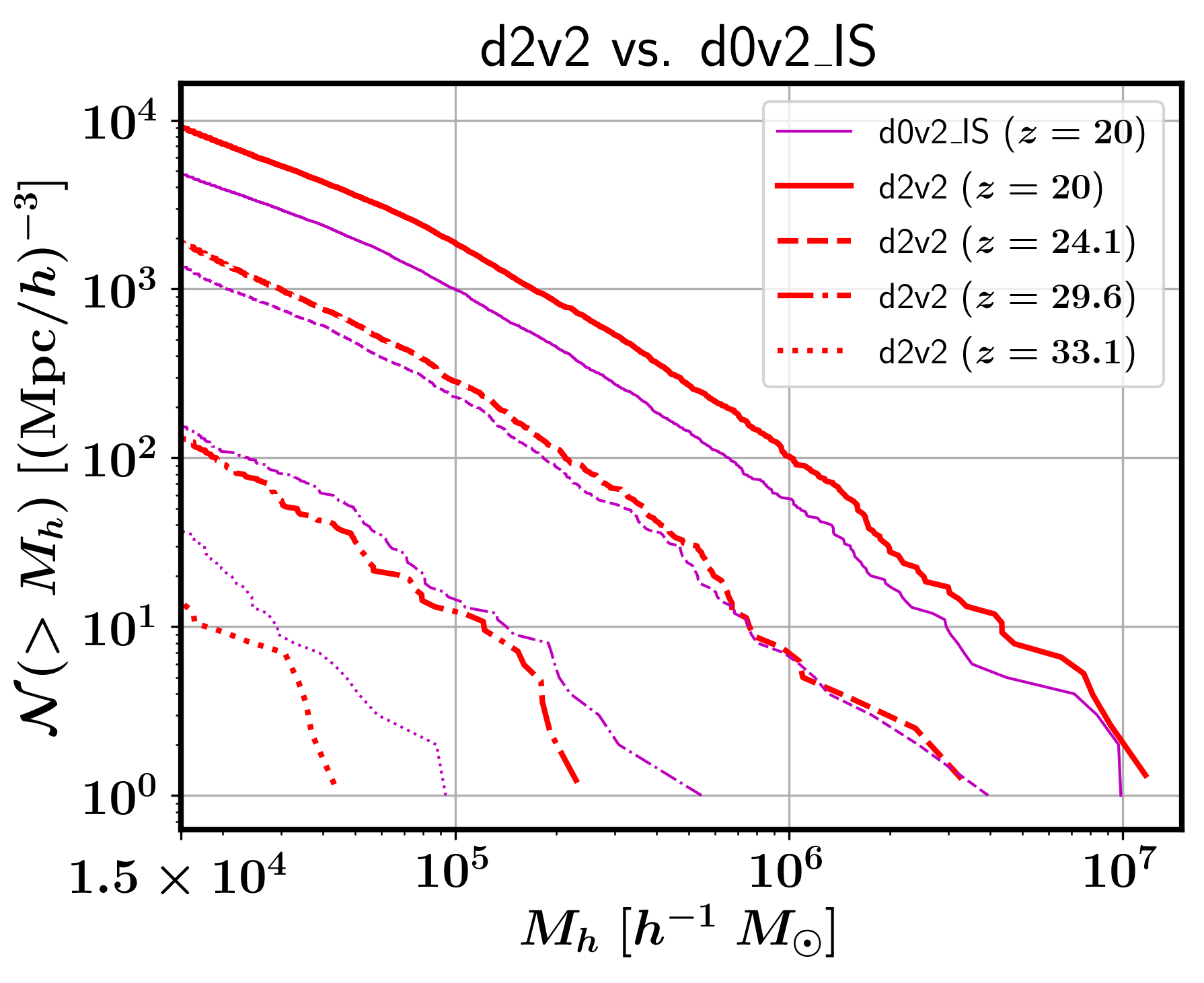}
  \caption{Left: accumulated halo mass function at $z=20$ as solid lines for d0v0 (black), d0v2 (cyan), d0v2\_BTD (blue), d2v2 (red), and d0v2\_IS (magenta). Right: accumulated mass function compared between d2v2 (red) and d0v2\_IS (magenta) for $z=20$ (solid), 24.1 (dashed), 29.6 (dotted-dashed), and 33.1 (dotted).}
  \label{fig:MF}
  \end{center}
\end{figure*}

\subsection{Streaming Effect with BTD} \label{sec:BTD}

\begin{figure*}
  \begin{center}
    \includegraphics[scale=0.45]{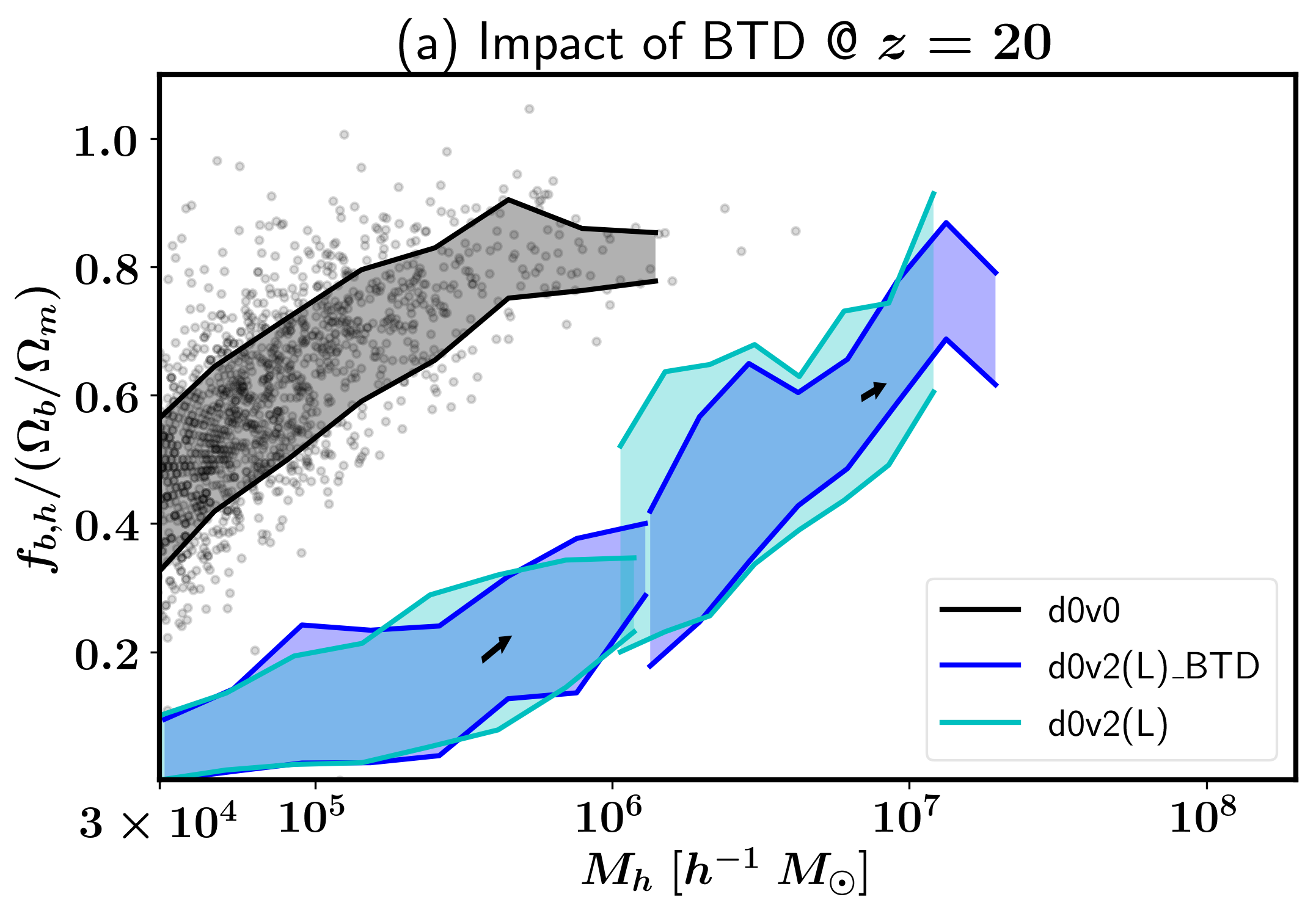}
    \includegraphics[scale=0.45]{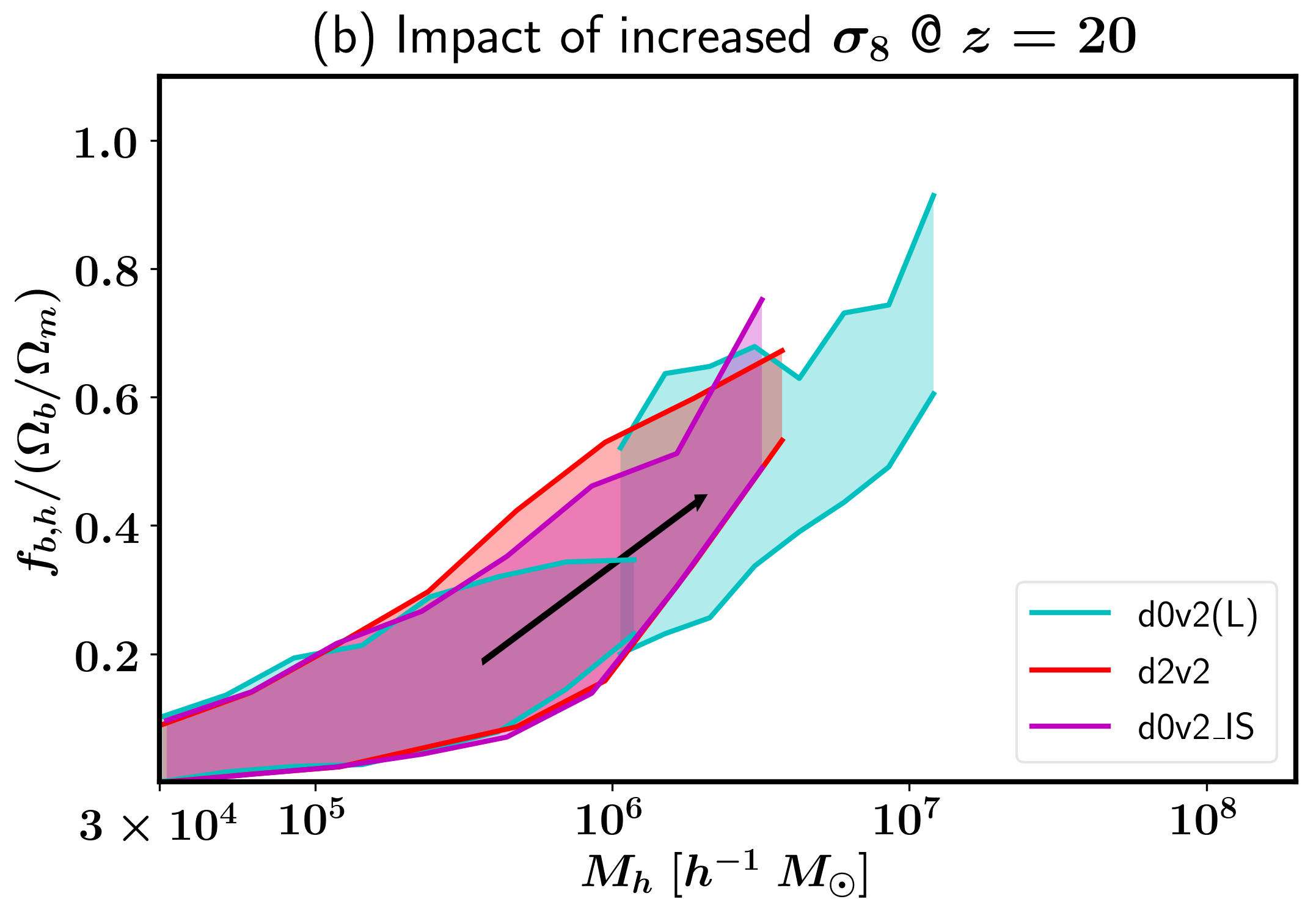}
\caption{Baryonic mass fraction of halos as a function of halo mass. The shade covers between 16th and 84th percentiles of the binned distribution of $f_{b,h}$. In the left panel, the results of d0v0 (black), d0v2 (cyan), d0v2L (cyan), d0v2\_BTD (blue), and d0v2L\_BTD (blue) at $z=20$ are compared. In the right panel, the results of d0v2 (cyan), d0v2L (cyan), d0v2\_IS (magenta), and d2v2 (red) are compared. Note that we use blue for both d0v2\_BTD and d0v2L\_BTD and cyan for both d0v2 and d0v2L. They can be distinguished from the mass range: d0v2 and d0v2\_BTD cover $M\lesssim 10^6~h^{-1}~ M_\odot$ while d0v2L and d0v2L\_BTD cover $M\gtrsim 10^6~h^{-1}~ M_\odot$. To illustrate the actual distribution of $f_{b,h}$, individual galaxies are shown as dots for d0v0 in the left panel. The tail and head of the two arrows in the left panel connects the averages for the 100 most massive halos in d0v2(L) and d0v2(L)\_BTD, respectively. Similarly, the arrow in the right panels connects the averages of 100 most massive halos in d0v2 and d2v2.}
  \label{fig:fb}
  \end{center}
\end{figure*}

We first examine the BTD approximation by comparing the $z=20$ snapshots of the no-streaming case (d0v0), the cases with streaming motion applied in the approximate way (d0v2\_BTD and d0v2L\_BTD), and the cases with the correctly implemented streaming motion (d0v2 and d0v2L), where all the cases except d0v0 have a streaming velocity of $56[z/1000]$ km/s. The gas particle maps are shown in Figure~\ref{fig:ptl_map}, the gas and dark matter density power spectra are shown for the simulations in $1~h^{-1}\rm{Mpc}$ boxes (d0v0, d0v2\_BTD, and d0v2) in the left panel of Figure~\ref{fig:PS}, the accumulated halo mass functions are shown in Figure~\ref{fig:MF}, and the baryonic mass fraction as a function of halo mass is shown in Figure~\ref{fig:fb}.

\subsubsection{Gas Density Fluctuation Amplitude}

The gas density maps in Figure~\ref{fig:ptl_map} visualize the well-known smoothing effect of the streaming motion: the density field appears much smoother in both d0v2 and d0v2\_BTD than in d0v0. Here, it is expected that the smoothing effect is underestimated in d0v2\_BTD compared to in d0v2 since the BTD assumption ignores the streaming effect in density taking place between the decoupling of baryons from photons ($z\approx1000$) and the beginning of the simulation ($z_i=200$ in this work). The difference in the fluctuation amplitude between in d0v2 and in d0v2\_BTD, which is not evident in the particle map, appears clearly in the comparison of gas density power spectra at $z=20$ (left panel of Fig.~\ref{fig:PS}), where the gas density power spectrum of d0v2\_BTD is up to $40\%$ larger than that of d0v2. The gas and dark matter density power spectra diverge above $k\sim30~h~{\rm Mpc}^{-1}$ in d0v2 while they diverge above $k\sim100~h~{\rm Mpc}^{-1}$ in d0v2\_BTD. The difference peaks at $k\approx 100~h~{\rm Mpc}^{-1}$ and decays toward high-$k$ until it vanishes at $k\approx 500~h~{\rm Mpc}^{-1}$. Clearly, the BTD approximation partially misses the streaming effect. 

We find that the difference in the shape of the gas density power spectrum between d0v2 and d0v2\_BTD can be explained from the distance covered by the streaming motion. The distance covered between $z_i$ and $z_f$ is given by
\bea
&&d_{cb}(z_i,z_f)=\int^{z_i}_{z_f} a^{-1} V_{cb}(z)  \frac{dt}{dz} dz \nonumber \\
&&\approx 2.1\times 10^{-3}\left[z_i^{0.5}-z_f^{0.5} \right]\left[\frac{V_{cb,1000}}{56~\rm km/s}\right]~h^{-1}~{\rm Mpc},
\eea
where we approximated the cosmic expansion rate to $a\approx z^{-1}$. In d0v2, the streaming velocity is accounted for from the decoupling of gas from the cosmic microwave background at $z_i\approx 1000$ giving $d_{cb}(z_i=1000,z_f=20)=5.8\times10^{-2}~h~{\rm Mpc}^{-1}$ and the gas density power spectrum in d0v2 hence deviates from the dark matter spectrum from roughly half the wavenumber corresponding to this distance $k_{cb}/2\equiv \pi/d_{cb}=5.5\times 10^1~h~\rm{Mpc}^{-1}$. In d0v2\_BTD, the streaming motion is accounted from the initialization redshift $z_i=200$ giving $d_{cb}(z_i=200,z_f=20)=2.0\times10^{-2}~h~{\rm Mpc}^{-1}$ and $k_{cb}/2=1.5\times10^2~h~\rm{Mpc}^{-1}$, which is similar to where the gas density power spectrum of d0v2\_BTD starts to deviate from the dark matter spectrum. The streaming distance ($d_{cb}$) is seemingly the threshold scale below which gas density fluctuations are suppressed. 

At scales much shorter than the streaming distance, the gas freely streams through the dark matter potential well. In that regime, the initial fluctuations in gas density would be completely washed out rapidly and the BTD approximation would not make any difference at later times. This explains why the gas density power spectra of d0v2 and d0v2\_BTD converge at $k\gtrsim 500~h~\rm{Mpc}^{-1}$. 

\subsubsection{Halo Abundance and Dark Matter Density Fluctuations}

The halo mass function in d0v2\_BTD is slightly higher than in d0v2 (See Fig.~\ref{fig:MF}). For example, the number of halos with $M_h>3\times10^5 M_\odot$ in d0v2 is reduced to 45.1\% of that in d0v0 and to 52.1\% in d0v2\_BTD. This difference is more or less constant throughout the entire range of halo mass in the simulation ($\lesssim 3\times 10^6~M_\odot$). A similar difference can be seen from the comparison of dark matter power spectra (solid lines in the left panel of Fig.~\ref{fig:ptl_map}) at $k\gtrsim 100~h~{\rm Mpc}^{-1}$. Compared to the impact of the streaming velocity, the error caused by the BTD approximation seems to be small in the dark matter sector.

\subsubsection{Baryonic Fraction in the Halo}\label{sec:fb}
The suppression of the baryonic fraction in halo mass $f_{b,h}$ is often considered as the most direct impact of the streaming motion on the Pop III star formation. Hence, $f_{b,h}$ has been repeatedly modeled with the streaming motion by previous works, most of which were based on the BTD approximation. 

Interestingly, the relation between $f_{b,h}$ and $M_h$ is not significantly changed by the approximation despite the bias introduced in the other statistics discussed above. Figure~\ref{fig:fb} shows that $f_{b,h}$ as a function of $M_h$ in d0v2 agrees with that in d0v2\_BTD within the 1$\sigma$ uncertainty up to $10^6~h^{-1} M_\odot$. Similarly, $f_{b,h}$ in d0v2L agrees with that in d0v2L\_BTD up to $10^7~h^{-1} M_\odot$. 

We find that the BTD approximation overestimates both $M_h$ and $f_{b,h}$ in a way that the $f_{b,h}$-$M_h$ relation is unchanged. For example, the averages of $M_h$ and $f_{b,h}$ for the 100 most massive halos are $3.7\times10^5~h^{-1}M_\odot$ and $0.19$ in d0v2, and $4.23\times10^5~h^{-1}M_\odot$ and $0.21$ in d0v2\_BTD. Similarly, the average of $M_h$ and $f_{b,h}$ are $6.95\times10^6~h^{-1}M_\odot$ and $0.60$ in d0v2L, and $7.70\times10^6~h^{-1}M_\odot$ and $0.61$ in d0v2L\_BTD. These changes are described by black arrows in Figure~\ref{fig:fb}, which lie along the direction of the $f_{b,h}$-$M_h$ relation. 

\subsection{Growth of Structure with Increased $\sigma_8$} \label{sec:IS}

Here, we compare the true overdense case (d2v2) to the artificial case where we boosted the initial density/velocity fluctuations in a mean-density volume by raising $\sigma_8$ (d0v2\_IS). We take d0v2 as the fiducial case so that the three cases (d2v2, d0v2\_IS, and d0v2) mentioned have the same streaming velocity of $V_{cb,1000}=56$ km/s. Such a boost of $\sigma_8$ has been commonly used to assimilate an overdense environment inside a mean-density simulation box. Note that increasing $\sigma_8$ is solely intended to mimic the overdensity environment regardless of $V_{cb}$, because $\Delta$ and $V_{cb}$ are mutually independent (\citealt{2016ApJ...830...68A,2018ApJ...869...76A}). The amount of increase in $\sigma_8$ would be identical to that in the case of, e.g., d0v0\_IS, if one were to use this scheme to mimic the d2v0 case. 

The density power spectrum (right panel of Fig.~\ref{fig:PS}) and the halo mass function (left panel of Fig.~\ref{fig:MF}) show how much more the structures have evolved in d2v2 and d0v2\_IS compared to in d0v2 at $z=20$. Both the density power spectrum and mass function show that the structures have grown more in d2v2 than in d0v2\_IS at $z=20$. However, a halo function comparison at higher redshifts in the right panel of Figure~\ref{fig:MF} shows the opposite: d0v2\_IS has more halos than d2v2 does at $z\gtrsim 30$.

\begin{figure}
  \begin{center}
    \includegraphics[scale=0.5]{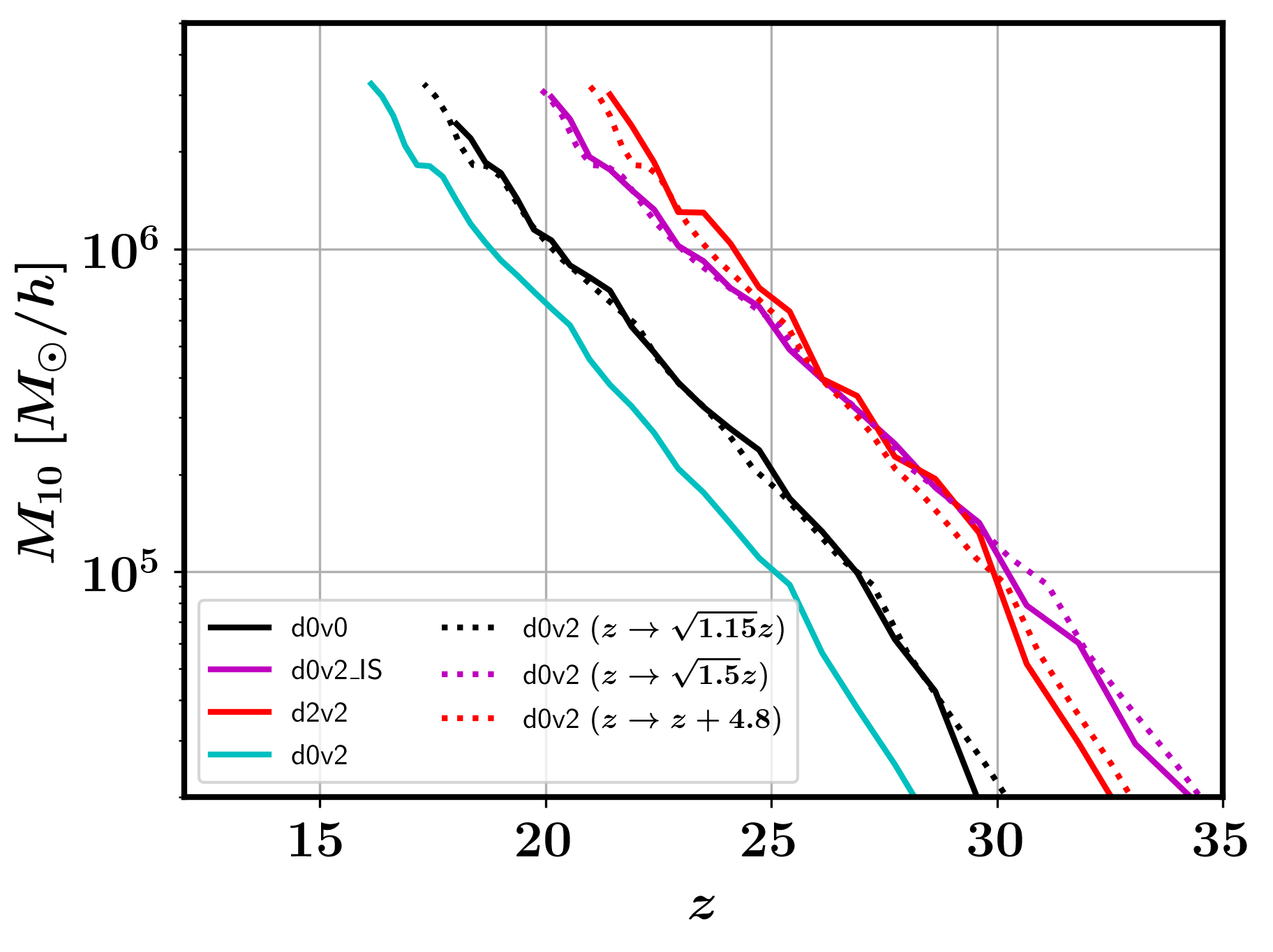}
  \caption{$M_{10}$ (Eq.~\ref{eq:M10}) as a function of redshift. The red, magenta, and cyan solid lines are the results of d2v2, d0v2\_IS, and d0v2, respectively. The magenta dashed line is the case where we transformed $M_{10}$ of d0v2 by $z\rightarrow \sqrt{1.5}z$ to match that of d0v2\_IS. Similarly, the black and redshift dotted lines are the results of transforming d0v2 with $z\rightarrow \sqrt{1.15}z$ and $z\rightarrow z+5$ to match d0v0 and d2v2, respectively.
   }
  \label{fig:Nm}
  \end{center}
\end{figure}

To compare the time evolution of halo mass function in a convenient manner, we define a mass $M_{10}$ in a way that the number of halos above that mass is fixed to a certain number density:
\bea \label{eq:M10}
\mathcal{N}(>M_{10})=10~(h^{-1}{\rm Mpc})^{-3}.
\eea
In Figure~\ref{fig:MF}, this would be the $x$-coordinate of the intersection of mass function and the second gray horizontal grid line from the bottom. Since the halo mass function grows monotonically in time, $M_{10}$ can be used as an indicator of how much the halos have grown in the simulation. We plot $M_{10}$  for d0v2, d2v2, and d0v2\_IS is shown in Figure~\ref{fig:Nm} as a function of redshift.

The redshift evolution of $M_{10}$ shows an interesting difference between d2v2 and d0v2\_IS. $M_{10}$ in d0v2\_IS (magenta solid) is larger than that of d2v2 (red solid) at $z\gtrsim 27$, but smaller at $z\lesssim 27$. This also agrees with the trend in the mass function comparison mentioned above (right panel of Fig.~\ref{fig:MF}).

The difference between d2v2 and d0v2\_IS in the time evolution of $M_{10}$ can be understood from how the structure formation is enhanced in those two cases. The structure growth in d0v2\_IS is given a head start at the beginning of the simulation and then proceeds just as fast as in the mean-density case later on. In contrast, the structure growth in d2v2 starts nearly the same as in the mean density case (d0v2) and is gradually accelerated by over time by a locally slower cosmic expansion rate.

Transforming $z$ to $\sqrt{1.5} z$ in $M_{10}(z)$ in d0v2 reproduces $M_{10}(z)$ in d0v2\_IS quite precisely (compare magenta solid and dotted lines in Fig.~\ref{fig:Nm}). This is explained by the growth rate of structure during the matter-dominated era: $P_{\delta\delta}(k)\propto a^2 \approx z^{-2}$. A factor of 1.5 increment in the initial density power spectrum results in the structure growth accelerated in the way $\sqrt{1.5}$ is multiplied to the redshift. We note that this effect is expected regardless of whether the streaming motion is present or not.

Interestingly, $M_{10}(z)$ in d0v0 is similar to transforming $z\rightarrow \sqrt{1.15} z$ in $M_{10}(z)$ in d0v2 (compare the black solid and dotted lines). According to the above finding, the impact of the 2-$\sigma$ level streaming velocity on halo mass growth history is similar to lowering $\sigma_8$ by $1.15$ at the initialization.

In the true overdense case (d2v2), the structure formation is accelerated by roughly a constant in redshift. $M_{10}$ in d2v2 is similar to transforming $z\rightarrow z+4.8$ in $M_{10}(z)$ in d0v2 throughout the range we explored ( $15\lesssim z \lesssim 35$; compare the red solid and dotted lines). This constant shift is smaller than the multiplicative shift in d0v2\_IS down to $z\approx 27$, but larger at lower redshifts.

The difference between d2v2 and d0v2\_IS indicates that self-consistent initial conditions are crucial in studying the effect of local overdensity on structure formation. Note that both the density power spectrum and the halo mass function are evaluated in the {\em global} comoving frame, and thus the results from the d2v2 simulation reflect the fact that the local patch has been detached and shrunken from the global comoving frame. Even though the boosted $\sigma_8$ of e.g. d0v2\_IS case can mimic the expedited formation of structures in overdense regions, this scheme cannot reproduce the density bias of halo clustering correctly because the simulation volume still has the same expansion as the global value. In terms of the peak-background split scheme \citep{Mo1996}, however, a boost of $\sigma_8$ (e.g., d0v2\_IS) only affects the linear density threshold for halo formation, but a locally collapsing patch (e.g., d2v2) affects both the halo-formation density threshold and the clustering scale of halos.

We find that the $f_{b,h}$-$M_h$ relation in d2v2 and in d0v2\_IS remains the same as in d0v2, d0v2 and d0v2\_BTD. That is, all the cases with $V_{cb,1000}=56$ km/s in this study show the same $f_{b,h}$-$M_h$ relation. Comparing d0v2 to d2v2 shows that the average baryonic fraction of the 100 most massive halos increasedsfrom $0.188$ to $0.433$ while the average halo mass is increases from $0.368$ to $1.92\times 10^6~h^{-1}M_\odot$ (see also the black arrow in the right panel of Figure~\ref{fig:fb}). This suggests that the relation depends only on the streaming velocity.

\section{Summary and Discussion} \label{sec:discussion}

Recently, a number of simulation studies have been performed in the context of assessing the impact of the baryon-dark matter streaming motion on the first collapsed objects. In this study, we have examined two approximations that are often made in those studies. One is the BTD approximation, which ignores the smoothing effect of the streaming motion in gas density before the initialization of simulation. The other is to boost the initial amplitude of density/velocity fluctuations to represent an overdense volume that forms the first collapsed objects.

The BTD approximation overestimates the gas density fluctuation amplitude by up to $\sim 40\%$ in the power spectrum at certain wavenumbers. The distance that the gas is shifted by the streaming motion is underestimated by roughly a factor of three as the streaming motion before the initialization of the simulation ($z_i=200$ in this study) is unaccounted for. As a result, the minimum wavenumber of the fluctuations suppressed by the streaming motion is overestimated by a similar factor. This results in the gas density power spectrum at $k\sim100~h~\rm{Mpc}^{-1}$ not being properly suppressed by the streaming motion. We note that this error is likely to be larger if a simulation is initialized at a lower redshift than in this study. 

On the other hand, the impact of the BTD approximation on dark matter structures is limited: the density power spectrum is overestimated by about 5\% at $k\gtrsim100~h~\rm{Mpc}^{-1}$ and the halo abundance is also increased by a similar fraction. The exaggeration in the total matter density fluctuation by the approximation is not severe since baryons do not dominate the gravitational growth of structures.

The baryonic mass fraction $f_{b,h}$ of a halo with its mass $M_h$, which is considered to be directly related to the Pop III star-formation is {\em not} significantly affected by the BTD approximation, justifying a number of previous works based on this approximation \citep[e.g.,][]{2017Sci...357.1375H,2019MNRAS.484.3510S}. The approximation overestimates both $M_h$ and $f_{b,h}$, but the relation between the two quantities remains unchanged. Moreover, the same relation holds for the halos in the overdense case (d2v2) and the increased $\sigma_8$ case (d0v2\_IS) as well, suggesting that $f_{b,h}$ depends only on $M_h$ and $V_{cb}$ and not on the gas density fluctuation amplitude. This can be explained by the fact that the gas density is not a dominant factor in the collapse of a halo: a dark matter clump first begins to collapse and the ambient gas is passively pulled in. $M_h$ would determine how much gas can potentially be pulled into the halo by gravity and $V_{cb}$ would determine how much of that gas is blown away.

Increasing $\sigma_8$ at the initialization of the simulation is a convenient way of studying a rare density peak that a Pop III star is expected to form. While the $f_{b,h}$-$M_h$ relation is not affected by the approximation as mentioned above, the growth history of halos is substantially different from in the true overdense case. Increasing $\sigma_8$ gives a head start in the structure growth while the growth in the true overdense volume is gradually accelerated over time. In that case, the structure growth and halo clustering are overestimated at the early time and underestimated at the late time, which would bias the growth history of the halo. For example, a $10^6~M_\odot$ minihalo at $z\approx30$ from simulations with an increased $\sigma_8$ of $1.2$ in \citet{2017Sci...357.1375H} will likely have a different mass growth history and halo-clustering scale than from in a true overdensity.

Interestingly, the way the halo growth is delayed by the streaming motion is in contrast to the impact of increasing $\sigma_8$. Presumably, the characteristic decay of the streaming velocity toward low redshift results in most of the suppression effect finishing much earlier than $z\sim30$, giving final structures similar to in the case of starting with a smaller density fluctuation amplitude at the initialization. In our simulation with $2\sigma$ streaming motion ($V_{cb,1000}$=56 km/s), the suppression in the halo mass function amounts to what we expect from lowering $\sigma_8$ by $13\%$ or transforming $z \rightarrow \sqrt{1.15}z$. This can be useful for modeling the impact of streaming on the global Pop III star formation rate \citep[e.g.,][]{2019PhRvD.100f3538M}. 

We note that our simulations do not include the chemical cooling needed to distinguish the cold component from the total minihalo gas. The streaming motion can reduce the cold fraction in the halo gas by shock heating, further reducing the chance of star formation in minihalos \citep{2019MNRAS.484.3510S} on top of the reduction in $f_{b,h}$. Given the small impact of the BTD approximation on halos, it is unlikely that including the chemical cooling would introduce a dramatic impact of the approximation on Pop III formation. It is, however, possible that increased $\sigma_8$ cases have some impact due to the biased growth history of halos. We aim to explore such issues in future studies to make more direct conclusions about the impact of the streaming motion on Pop III star formation.

\section*{Acknowledgement}
 The authors thank the anonymous reviewer for his/her helpful comments. This work was supported by the World Premier International Research Center Initiative (WPI), MEXT, Japan. HP was supported by JSPS KAKENHI Grant Number 19K23455. KA was supported by NRF-2016R1D1A1B04935414 and NRF-2016R1A5A1013277, and appreciates APCTP and IPMU for their hospitality during completion of this work. SH was supported by JSPS KAKENHI Grant Number 18J01296. Numerical computations were carried out on Cray XC50 and PC cluster at Center for Computational Astrophysics, National Astronomical Observatory of Japan.

\end{CJK}
\bibliographystyle{apj}
\bibliography{reference}
\end{document}